\documentclass[,final]{aipproc}

\layoutstyle{6x9}

\newcommand{\be}{\begin{eqnarray}}
\newcommand{\ee}{\end{eqnarray}}

\begin{document}

\title{The $e^+e^-\rightarrow J/\psi D \bar D$, $J/\psi D\bar D^*$ reactions with dynamically generated resonances}

\keywords      {Dynamically generated resonances, $X(3872)$}

\classification{}

\author{D. Gamermann}{
  address={Departamento de F\'isica Te\'orica and IFIC, Centro Mixto
Universidad de Valencia-CSIC,\\ Institutos de Investigaci\'on de
Paterna, Aptdo. 22085, 46071, Valencia, Spain}
}

\author{E. Oset}{
  address={Departamento de F\'isica Te\'orica and IFIC, Centro Mixto
Universidad de Valencia-CSIC,\\ Institutos de Investigaci\'on de
Paterna, Aptdo. 22085, 46071, Valencia, Spain}
}

\begin{abstract}
In two recent reactions by Belle producing $D\bar D$ and $D\bar D^*$ meson pairs, peaks above threshold have been measured in the differential cross sections, possibly indicating new resonances in these channels. We want to study such reactions from the point of view that the $D$ meson pairs are produced from already known or predicted resonances below threshold. Our study shows that the peak in the $D\bar D^*$ production is not likely to be caused by the $X(3872)$ resonance, but the peak seen in $D\bar D$ invariant mass can be well described if the $D\bar D$ pair comes from the already predicted scalar $X(3700)$ resonance.
\end{abstract}

\maketitle

\section{I. Introduction}

Belle collaboration has measured reactions producing $D$ mesons together with a $J/\psi$ from $e^+e^-$ collisions \cite{belle}. The invariant mass distributions for $D\bar D$ and $D\bar D^*$ in these reactions peak around 70 MeV above threshold. This observation lead to the claim of new resonances. In a previous paper \cite{belle2} Belle has studied the $D\bar D^*$ invariant mass distribution in $B$ decays ($B\rightarrow KD\bar D^*$), in this case a peak was observed only a few MeV above threshold and it was first suggested that this peak could be indicative of the existence of a new $X(3875)$ resonance.

This second experiment has been analyzed in several theoretical works. In \cite{hanh} it is concluded that the peak seen in \cite{belle2} may be caused by the $X(3872)$ if it is a virtual state. A latter work \cite{nml} shows that this conclusion may be changed if one takes into account explicitly the $D^*$ width. A different approach is suggested in \cite{meuax}, where the $X(3872)$ is generated dynamically by the interaction of pseudoscalar mesons with vector ones.

The idea of generating dynamically resonances in order to describe the newly found charmed states is not new and had already been exploited in \cite{lutz1,lutz2,guo1,guo2}. We have extended the ideas of these works including the hidden charm sector in \cite{meuax,meusca}.

This work presents the results of \cite{meuhid} in which the peaks seen in \cite{belle} are interpreted as caused by some of the hidden charm subthreshold resonances dynamically generated in \cite{meuax,meusca}.

\section{II. Framework}

\begin{figure}[h]
\includegraphics[width=6cm]{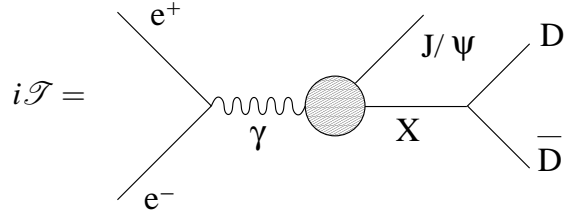} 
\put(-210,40){$i{\cal T}=$}
\caption{Feynman diagram for the process $e^+e^-\rightarrow J/\psi D\bar D$} \label{fig1}
\end{figure}

We assume that the $D$ mesons in the reaction $e^+e^-\rightarrow J/\psi D\bar D$ come from a resonance. In this case the reaction can be described by the diagram in figure \ref{fig1} and close to the $D\bar D$ threshold the amplitude depends strongly only on the $D\bar D$ invariant mass.

This amplitude can be written in terms of the $X$ propagator with all other parts factorized as a constant $C$:

\begin{eqnarray}
{\cal T}&=&C {1\over M_{inv}^2(D\bar D)-M_X^2+i\Gamma_X M_X}, \label{eq1}
\end{eqnarray}
if we describe the $X$ resonance as a Breit-Wigner type.

The cross section can be calculated from the amplitude by integrating over the phase space of the three particles in the final state. Since $\cal T$ depends only on the $D\bar D$ invariant mass, one can evaluate the differential cross section that takes the form:

\begin{eqnarray}
{d\sigma\over dM_{inv}(D\bar D)}&=&{1\over(2\pi)^3}{m_e^2\over s\sqrt{s}} |\overrightarrow k| |\overrightarrow p| |{\cal T}|^2, \label{dcross}
\end{eqnarray}
where $s$ is the center of mass energy of the electron positron pair squared and $|\overrightarrow k|$ and $|\overrightarrow p|$ are given by the relative momentum of the $D$ mesons and the relative momentum of the $J/\psi$ and the $D$ meson system in the rest frame of each system respectively.

Now we are going to describe the model for generating dynamically the resonances. First a field is written for all mesons belonging to a 15-plet of $SU(4)$, one field, $\Phi$, for the pseudoscalars and another, ${\cal V}_\mu$ for the vector mesons. Explicit expressions for these fields can be found in \cite{meuax,meusca}.

The Lagrangians are built by coupling two currents of the meson fields. The Lagrangians built this way are $SU(4)$ invariant, but since this is not an exact symmetry in Nature, one breaks it by suppressing the terms of these Lagrangians where the underlying interaction is driven by the exchange of a heavy vector mesons \cite{hofmann,angels}.

One collects the tree level amplitudes for meson-meson scattering, projected in s-wave, for each couple channel space, into a matrix. This matrix is used as input to solve the Bathe-Salpeter equation that in the on-shell formalism of \cite{noverd,oller}, assumes an algebraic form:

\begin{eqnarray}
T&=&(\hat 1-VG)^{-1}V \textrm{, for scalars} \label{bseq1} \\
T&=&-({\hat 1} + V{\hat G})^{-1}V \overrightarrow{\epsilon}.\overrightarrow{\epsilon}' \textrm{, for axials} \label{bseq2}
\end{eqnarray}

The loop function, $G_{ii}$, has its imaginary part fixed in order to ensure unitarity of the T-matrix. Resonances are identified as poles in the second Riemann sheet of the T-matrix.

Within this model the $D_{s0}^*(2317)$, $D_{s1}(2460)$ and $X(3872)$ are well reproduced. Moreover interesting predictions are a heavy hidden charm scalar with mass around the $D\bar D$ threshold and an axial state with negative C-parity, almost degenerated in mass with the $X(3872)$. Also broad open charm states with exotic quantum numbers appear. These results are, though, less stable and contrast with previous works done where states with the same quantum numbers are narrower \cite{lutz1,lutz2,guo1,guo2}.

The way we are going to apply this model to describe the reaction of figure \ref{fig1} is to substitute the Breit-Wigner amplitude of eq. (\ref{eq1}) by the $D\bar D$ T-matrix calculated from eq. (\ref{bseq1}). For the reaction $e^+e^-\rightarrow J/\psi D\bar D^*$ everything is done analogously using eq. (\ref{bseq2}) for the T-matrix.

\section{III. Results}

The $e^+e^-$ process measured by Belle \cite{belle} has a center of mass energy of $\sqrt{s}=10.6$ GeV. The $M_{inv}(D\bar D^{(*)})$ is measured from threshold up to 5.0 GeV. Since our model is reliable for energies within a few hundreds of MeV around the thresholds, we compare our model with belle's data for the points below 4.2 GeV. The invariant mass distributions are measured in counts per bin. To simulate the bins we integrate our theoretical curves in bins of the same size as experiment and normalize so that the total integral of our curve matches the total number of events in the experiment.

Finally we compare the data with our predictions by a standard $\chi^2$ test.

The model has one free parameter, which is the subtraction constant in the loop function, $\alpha_H$. We show results for different values of $\alpha_H$ which correspond to a natural size. For the case of two $D$ mesons in a loop, for example, the value $\alpha_H$=-1.3 correspond to a cut off with $q_{max}=$ 850 MeV. We present results for different values of $\alpha_H$ together with the corresponding pole position.

\begin{figure}[t]
\begin{tabular}{cc}
\includegraphics[width=5cm,angle=-0]{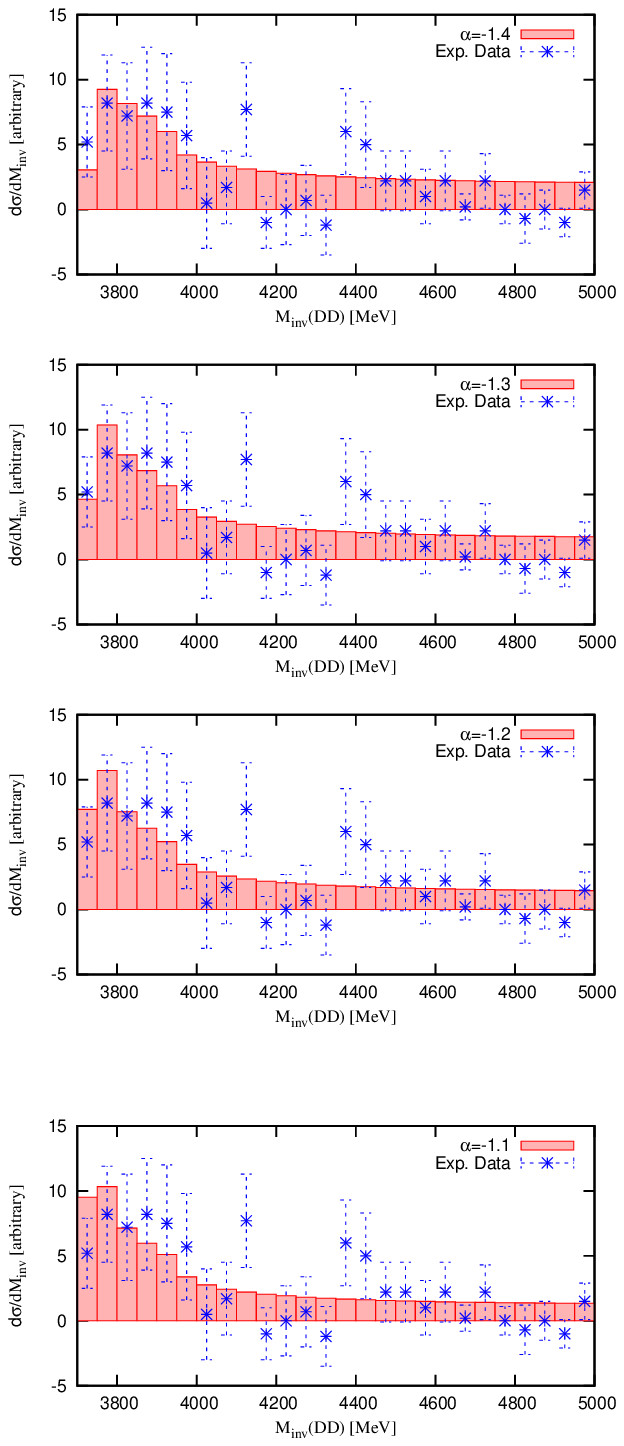} &
\includegraphics[width=5cm,angle=-0]{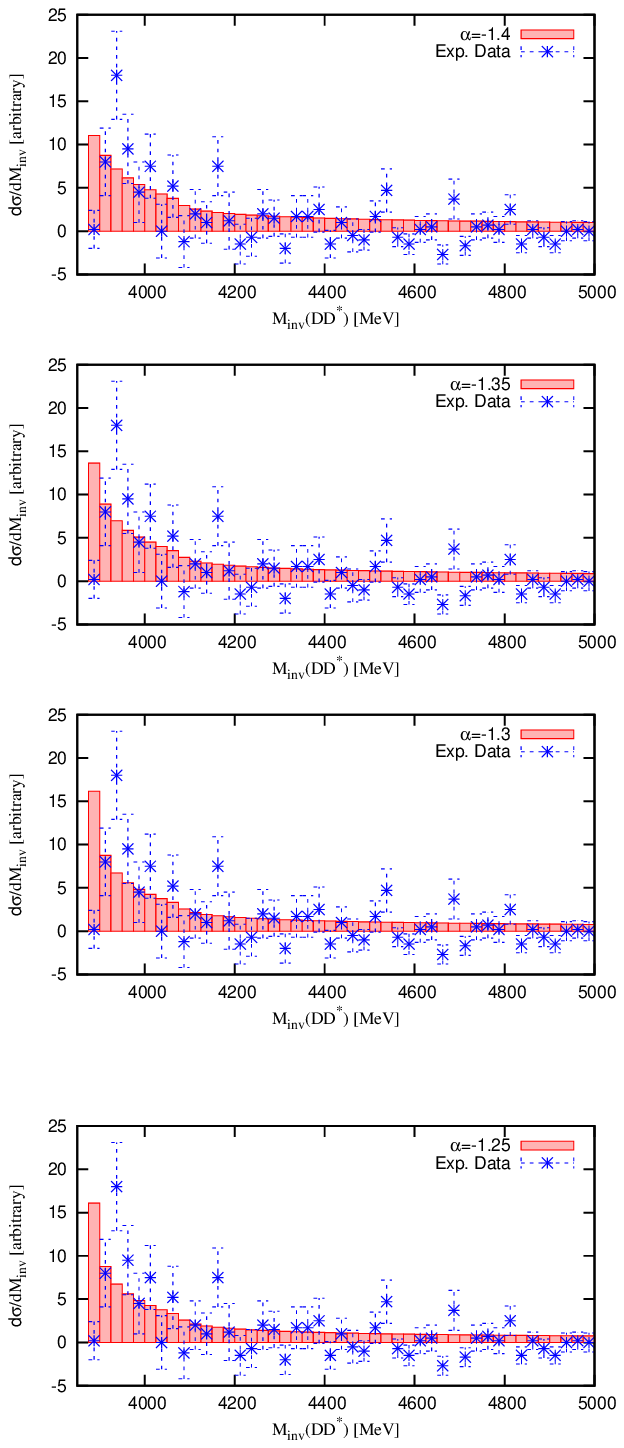} 
\end{tabular}
\caption{Left: Theoretical histograms compared with data from \cite{belle} for $D\bar D$ invariant mass distribution. Right: Theoretical histograms compared with data from \cite{belle} for $D\bar D^*$ invariant mass distribution.} \label{fig2}
\end{figure}

In table \ref{tab1} we show results, in the scalar sector, for the value of $\chi^2$ calculated with the data from Belle, for all points below 4.2 GeV in the $J/\psi D\bar D$ and $J/\psi D\bar D^*$ production. Fig. \ref{fig2} shows plots of our theoretical histograms compared with the experimental data \cite{belle}.

\begin{table}
\caption{Results of $M_X$ and $\chi^2$ for different values of $\alpha_H$.} \label{tab1}
\begin{tabular}{c|cc||c|cc}
\hline
\multicolumn{3}{c||}{$X(3700)$} & \multicolumn{3}{c}{$X(3872)$} \\
\hline
$\alpha_H$ & $M_X$(MeV) & $\chi^2 \over d.o.f$ &$\alpha_H$ & $M_X$(MeV) & $\chi^2 \over d.o.f$\\
\hline
\hline
-1.4   & 3701.93-i0.08 & 0.96&-1.40   &3865.09-i0.00& 3.36 \\
-1.3   & 3718.93-i0.06 & 0.85&-1.35   &3870.07-i0.00& 4.54  \\
-1.2  & 3728.12-i0.03 & 0.92&-1.30  & 3872.67-i0.00 & 5.96 \\
-1.1   & Cusp & 1.11&-1.25   & Cusp & 5.92 \\
\hline
\end{tabular}
\end{table}

One can see that the values of $\chi^2$ are around 1 for the $X(3700)$, indicating a good fit to the data. Although the large experimental errors are in part responsible for this good fit the data is compatible with the existence of the scalar resonance below the $D\bar D$ threshold.

In the case with a $D\bar D^*$ pair in the final state we have to apply our model for the interaction of pseudoscalars with vector mesons, in this case the resonances generated are axials, and the $X$ in figure \ref{fig1} should be identified with the $X(3872)$ generated dynamically in our model. Table \ref{tab1} also shows results for $M_X$ and $\chi^2$ for different values of $\alpha_H$ for this sector. Since the state $X(3872)$ is known and has a rather precise mass, we have chosen a smaller range to vary the parameter $\alpha$.

In this case the $\chi^2$ obtained is in all cases bigger than 3, clearly indicating a poor fit to the data.

\begin{figure}[h]
\begin{tabular}{cc}
\includegraphics[width=5cm,angle=-0]{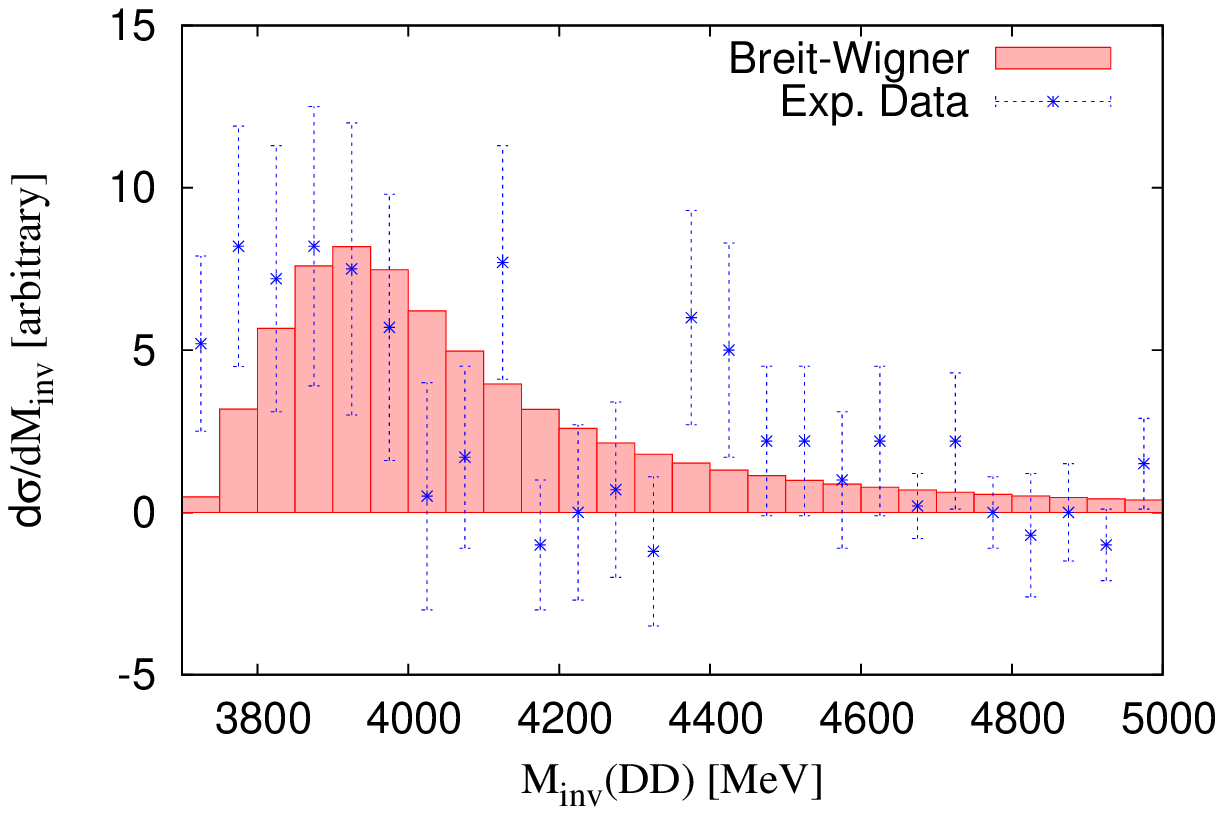} 
&
\includegraphics[width=5cm,angle=-0]{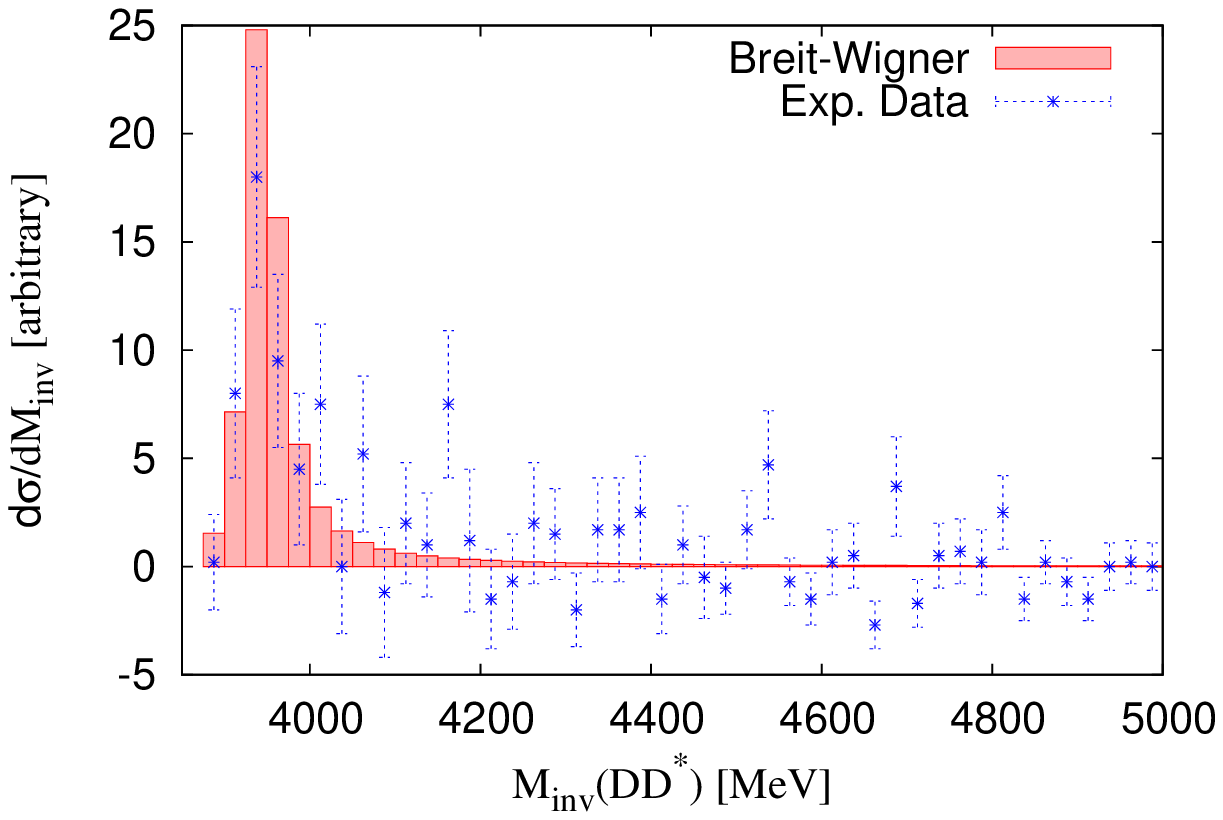} 
\end{tabular}
\caption{Left: Histograms calculated with Breit-Wigner resonance with mass $M_X$=3940 MeV compared to data. Right: Histograms calculated with Breit-Wigner resonance with mass $M_X$=3880 MeV compared to data.} \label{fig4}
\end{figure}

In the experimental paper the peaks have been fitted with Breit-Wigner type resonances \cite{belle}, suggesting two new States. We have also performed such a fit in order to compare the results and have more meaningful conclusions. We take the same Breit-Wigner parameters suggested in the experimental paper. The scalar resonance with $M_X$=3878 MeV and $\Gamma_X$=347 MeV and the axial one with $M_X$= 3942 MeV and $\Gamma_X$= 37 MeV. We show the results obtained by fitting a Breit-Wigner form from eq. (\ref{eq1}) in $\cal T$ of eq. (\ref{dcross}) in figure \ref{fig4}. Additionally we calculate $\chi^2$ and find $\chi^2/d.o.f$=2.10 for the $D\bar D$ distribution and $\chi^2/d.o.f$=1.34 for the $D\bar D^*$ distribution. The value of $\chi^2$ for the $D\bar D$ distribution can be improved if we take different parameters for the Breit-Wigner resonance. Taking for the fit $M_X$=3750 MeV and $\Gamma_X$=250 MeV we obtain a value of $\chi^2/d.o.f$=1.12, still slightly bigger than those obtained in our previous analysis. The value of $\chi^2$ for the $D\bar D^*$ distribution is undoubtedly better in the case of a Breit-Wigner fit that in our analysis assuming the $X(3872)$ resonance as the $X$ in the mechanism of fig. \ref{fig1}.

\section{IV. Summary}

We have analyzed a recent experiment from Belle \cite{belle} where peaks are observed in the $D$ meson pair invariant mass suggesting new states with hidden charm. We have calculated the differential cross sections for these measurements from the point of view that the $D$ meson pairs are produced by dynamically generated resonances below threshold.

Our theoretical model describes very well the data for the $D\bar D$ invariant mass distribution as being caused by a dynamically generated scalar resonance below $D\bar D$ threshold. On the other hand, the $\chi^2$ values obtained in the fit of our theoretical curves for the $D\bar D^*$ invariant mass distribution are too big, favoring thus the interpretation of a new state with mass around 3940 MeV.

Although we have a good description of the scalar resonance, the large experimental uncertainties do not discard other possibilities. New experiments to search for this predicted scalar state are most welcome.


\begin{thebibliography}{9}

\bibitem{belle}
   K.~Abe {\it et al.}  [Belle Collaboration],
  arXiv:0708.3812 [hep-ex].

\bibitem{belle2}
  G.~Gokhroo {\it et al.},
  Phys.\ Rev.\ Lett.\  {\bf 97}, 162002 (2006)
  [arXiv:hep-ex/0606055].

\bibitem{hanh}
  C.~Hanhart, Yu.~S.~Kalashnikova, A.~E.~Kudryavtsev and A.~V.~Nefediev,
  arXiv:0704.0605 [hep-ph].

\bibitem{nml}
  E.~Braaten and M.~Lu,
  arXiv:0709.2697 [hep-ph].

\bibitem{meuax}
 D.~Gamermann and E.~Oset,
  Eur.\ Phys.\ J.\  A {\bf 33}, 119 (2007)
  [arXiv:0704.2314 [hep-ph]].

\bibitem{lutz1}
 E.~E.~Kolomeitsev and M.~F.~M.~Lutz,
  Phys.\ Lett.\ B {\bf 582} (2004) 39
  [arXiv:hep-ph/0307133].

\bibitem{lutz2}
 J.~Hofmann and M.~F.~M.~Lutz,
  Nucl.\ Phys.\ A {\bf 733}, 142 (2004)
  [arXiv:hep-ph/0308263].

\bibitem{guo1}
 F.~K.~Guo, P.~N.~Shen and H.~C.~Chiang,
  Phys.\ Lett.\  B {\bf 647}, 133 (2007)
  [arXiv:hep-ph/0610008].

\bibitem{guo2}
 F.~K.~Guo, P.~N.~Shen, H.~C.~Chiang and R.~G.~Ping,
  Phys.\ Lett.\ B {\bf 641} (2006) 278
  [arXiv:hep-ph/0603072].

\bibitem{meusca}
  D.~Gamermann, E.~Oset, D.~Strottman and M.~J.~Vicente Vacas,
  Phys.\ Rev.\  D {\bf 76}, 074016 (2007)
  [arXiv:hep-ph/0612179].

\bibitem{meuhid}
  D.~Gamermann and E.~Oset,
  arXiv:0712.1758 [hep-ph].

\bibitem{hofmann}
  J.~Hofmann and M.~F.~M.~Lutz,
  Nucl.\ Phys.\  A {\bf 763}, 90 (2005)
  [arXiv:hep-ph/0507071].

\bibitem{angels}
  T.~Mizutani and A.~Ramos,
  Phys.\ Rev.\  C {\bf 74}, 065201 (2006)
  [arXiv:hep-ph/0607257].

\bibitem{skyrme}
  H.~Walliser,
  Nucl.\ Phys.\  A {\bf 548}, 649 (1992).

\bibitem{noverd}
  J.~A.~Oller and E.~Oset,
  Phys.\ Rev.\  D {\bf 60}, 074023 (1999)
  [arXiv:hep-ph/9809337].

\bibitem{oller}
  J.~A.~Oller and U.~G.~Meissner,
  Phys.\ Lett.\  B {\bf 500}, 263 (2001)
  [arXiv:hep-ph/0011146].



\end{thebibliography}
\end{document}